\documentclass[sigconf]{acmart}

\usepackage{booktabs} % For formal tables
\usepackage{kotex}
\usepackage[ruled,lined,linesnumbered,noend]{algorithm2e}
\usepackage{ctable}
\usepackage{balance}
\setcopyright{rightsretained}
\newcommand{\eg}{{\textit{e.g.}}}
\newcommand{\ie}{{\textit{i.e.}}}

\SetKwProg{Fn}{Function}{}{}

\acmDOI{10.475/123_4}

\acmISBN{123-4567-24-567/08/06}

\acmConference[IFUP'18]{Workshop on Multi-dimensional Information Fusion for User Modeling and Personalization}{Feb 2018}{Los Angeles, USA} 
\acmYear{2018}
\copyrightyear{2018}

\begin{document}
\title{Reinforcement Learning based Recommender System\\using Biclustering Technique}
% \titlenote{Produces the permission block, and copyright information}
% \subtitle{Extended Abstract}
% \subtitlenote{The full version of the author's guide is available as
%   \texttt{acmart.pdf} document}

\author{Sungwoon Choi}
% \authornote{Dr.~Trovato insisted his name be first.}
%\authornote{Sungwoon Choi is also a member of the Korea University.}
\affiliation{%
  \institution{Seoul National University}
  \institution{Korea University}
%   \city{Seoul, Korea}
%   \postcode{08826}
}
\email{nebulach23@gmail.com}

\author{Heonseok Ha}
\affiliation{%
  \institution{Seoul National University}
%   \city{Seoul, Korea}
%   \postcode{08826}
}
\email{heonseok.ha@gmail.com}

\author{Uiwon Hwang}
% \authornote{This author is the
%   one who did all the really hard work.}
\affiliation{%
  \institution{Seoul National University}
%   \city{Seoul, Korea}
%   \postcode{08826}
}
\email{uiwon.hwang@snu.ac.kr}

\author{Chanju Kim}
\affiliation{%
\institution{Clova AI Research}
  \institution{NAVER Corporation}
}
\email{chanju.kim@navercorp.com}

\author{Jung-Woo Ha}
\affiliation{
\institution{Clova AI Research}
  \institution{NAVER Corporation}}
\email{jungwoo.ha@navercorp.com}

\author{Sungroh Yoon}
\authornote{Corresponding author}
\affiliation{%
  \institution{Seoul National University}
%   \city{Seoul, Korea}  
%   \postcode{08826}
  }
\email{sryoon@snu.ac.kr}

\renewcommand{\shortauthors}{S. Choi et al.}

\begin{abstract}
A recommender system aims to recommend items that a user is interested in among many items. The need for the recommender system has been expanded by the information explosion. Various approaches have been suggested for providing meaningful recommendations to users. One of the proposed approaches is to consider a recommender system as a \textit{Markov decision process} (MDP) problem and try to solve it using reinforcement learning (RL). However, existing RL-based methods have an obvious drawback. To solve an MDP in a recommender system, they encountered a problem with the large number of discrete actions that bring RL to a larger class of problems. In this paper, we propose a novel RL-based recommender system. We formulate a recommender system as a gridworld game by using a biclustering technique that can reduce the state and action space significantly. Using biclustering not only reduces space but also improves the recommendation quality effectively handling the cold-start problem. In addition, our approach can provide users with some explanation why the system recommends certain items. Lastly, we examine the proposed algorithm on a real-world dataset and achieve a better performance than the widely used recommendation algorithm.
\end{abstract}


% We no longer use \terms command


\maketitle

\section{Introduction}
As the choice of users increases, the importance of recommender systems that assist in decision making is increasing day by day. Recommender systems are introduced in a variety of domains, and the performance of recommender systems is directly related to the interests of the company or individual. Previously, recommender systems have achieved great success with a method called collaborative filtering (CF). CF is one of the most popular techniques in the recommender system domain. The objective of CF is to make a personalized prediction about the preferences of users using the information about other users who have similar interests for items. 

    One disadvantage of CF is that it considers only one of the two dimensions (\ie, users or items), which often makes it difficult to detect important patterns that otherwise could be captured by considering both dimensions. In addition, the data matrix a typical recommender system has to handle is sparse and high-dimensional, because there are a large number of available items, many of which are never purchased or rated by the users. These two facts led to the developments of biclustering-based recommender systems, some of which have shown superior performance to conventional CF approaches~\cite{symeonidis2008nearest,zhang2014cold,leung2011clr,george2005scalable,alqadah2015biclustering}. Biclustering, also known as co-clustering~\cite{dhillon2001co}, two-way clustering~\cite{getz2000coupled}, and simultaneous clustering~\cite{jornsten2003simultaneous}, aims to find subsets of rows and columns of a given data matrix~\cite{cheng2000biclustering}. The big difference between clustering and biclustering is that clustering derives a global model, whereas biclustering produces a local model~\cite{madeira2004biclustering}.
    
Another disadvantage of CF is that it is static, therefore it is usually not possible to reflect a user's response in real time. Therefore, an MDP-based recommender system is proposed~\cite{shani2005mdp}. They use a discrete state MDP model to maximize the utility function that takes into account the future interactions with their users. In their work, they suggest the use of an $n$-gram predictive model for generating the initial MDP. They consider the actions of the MDP as a recommendation for an item. This leads to a large action space which makes it difficult to solve the MDP problem. 

In this paper, we propose a new recommendation algorithm using biclustering and RL. We reduce state and action space by using a biclustering technique which renders the MDP problem easy to solve. Using biclustering not only reduces space but also improves the recommendation quality of the cold start problem. Moreover, it can be explained to users why the system recommends certain items. 

The paper is structured as follows. In Section~\ref{sec:preliminary} we review the necessary background on MDP and RL. In Section~\ref{sec:problem_definition} we define the problem. In Section~\ref{sec:proposed_approach} we describe the proposed approach. Section~\ref{sec:experiments} provide an empirical evaluation of the actual recommender system based on the two Movielens datasets. We discuss the paper in Section~\ref{sec:dicussion} then we conclude the paper in Section~\ref{sec:conclusion}.

\section{Preliminary}\label{sec:preliminary}
\textbf{\textit{Markov Decision Processes :}} An MDP is a model for sequential stochastic decision problems~\cite{sutton1998reinforcement}. An MDP model is specified by a tuple of states, actions, a reward function, a transition function, and a discount factor. The agent stays in a particular state $s_t \in S$ for each discrete time step $t \in \{0,1,2,...\}$. After the choice of an action $a_t \in A$, the agent moves to the next state $s_{t+1}$ by calling a transition function $\textrm{T}(s_t,a_t)$. At the same time the agent receives a reward $r_{t}$ from the environment by reward function $\textrm{R}(s_t,a_t,s_{t+1})$. Based on a policy $\pi(s)$, the action $a$ is selected in a certain state $s$. MDP can be solved by RL. RL aims to find the optimal policy $\pi^*$ that maximizes the exptected cumulative reward $G$ which is called return. In RL, the optimal policy can be learned by a state-action value function $\textrm{Q}_{\pi}(s, a)$ which means the expected value of the return $G$ obtained from episodes starting from a certain state $s$ with the action $a$. $\textrm{Q}_{\pi}(s, a)$ can be expressed as follows:
\begin{align}
    \textrm{Q}_{\pi}(s, a) &= E_{\pi}\{G_t | s_t =s, a_t = a \}\\
    &= E_{\pi}\left \{  \sum_{k=0}^{\infty } \gamma^kr_{t+k}|s_t = s, a_t = a  \right \}
\end{align}
where $\gamma$ is a discount factor $(0 < \gamma \leq 1)$. 

\section{Problem Definition}\label{sec:problem_definition}
\begin{figure}[t]
    	\begin{center}
    		\includegraphics[width=2.8in]{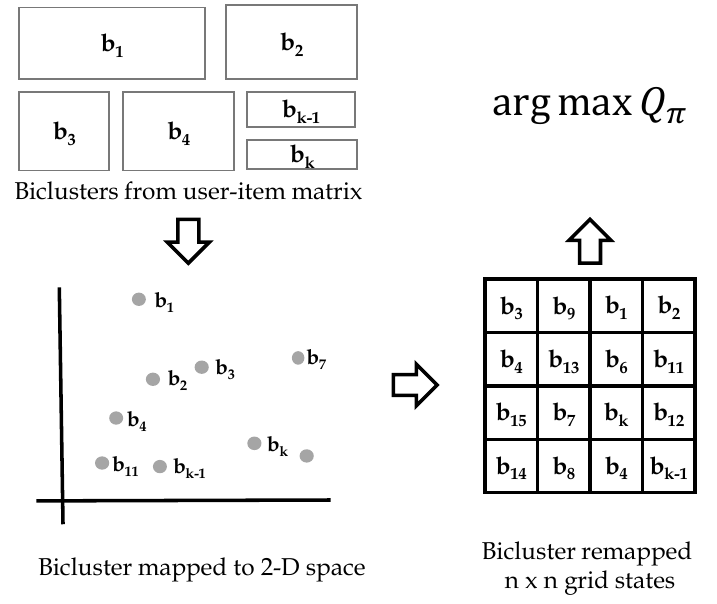}
    	\end{center}
    	\caption{\small Overview of the proposed method}
    	\label{fig:overview}
    \end{figure}

	We consider a recommender system as an MDP problem that can be formalized in a gridworld. Figure \ref{fig:overview} describes the overview of the formalization. A gridworld is a 2D environment in which an agent can move in four directons at a time. Typically, the goal in a gridworld is that the agent navigates to some location by maximizing the return. In our case, the agent and the state are considered as a user and a group of items, respectively. User movement in the gridworld means getting new recommendations from the group of items. Moreover, the reward can be considered as a user's satisfaction for the recommended items.
	At first, we need to be specify the environment of the MDP. In this paper, we assume that we have obtained $n^2$ biclusters from the user and item matrix $B=(U,I)$. We describe the environment in more detail below: 

\textbf{State Space $S$} : Gridworld has $n \times n = n^2$ distinct states. Each state $s=(U, I)$ includes a user set $U$ and an item set $I$ which are obtained from biclustering. The start state can be any state. Outside of the gridworld cannot be moved to.

\textbf{Action Space $A$} : The agent can choose from up to four actions to move around: up, down, left, right.

\textbf{Transition Function $\textrm{T}(s_t, a_t)$} : Gridworld is deterministic.

\textbf{Reward Function $\textrm{R}(s_t,a_t,s_{t+1})$} : A reward $r_t$ is also deterministic and is determined by the proposed reward function as follows:
\begin{align}
\textrm{R}(s_t,a_t,s_{t+1}) &= \textrm{Jaccard\_Distance}\-(U_{s_t}, U_{s_{t+1}}) \\ &= \frac{|U_{s_t} \cap U_{s_{t+1}}|}{|U_{s_t} \cup U_{s_{t+1}}|}.
\end{align}
The agent receives between 0 to 1 reward through the calculation of the Jaccard distance with the user vectors of two states $s_t, s_{t+1}$. In this environment, reward is deterministic function of state-action pair. As the two states have more same users, the reward approaches to 1. The similarity of the two item vectors of the states is not considered as a reward, because we do not want to recommend only a small number of items when moving the state. 
    
\section{Proposed Approach}\label{sec:proposed_approach}

	 The proposed method is composed of four parts: constructing the states, learning the $Q$-Function, generating recommendations, and updating the model online.
       
\subsection{Constructing the States}\label{sec:discovering_mapping}
    
    In the next step, each state is mapped to one of $n^2$ biclusters, so that each state has an item set and a user set. The mapping is performed based on the distance between the user vector of the bicluster and the states of the gridworld that can be considered as a two-dimensional (2D) euclidean space. However, the user vector of the bicluster is not 2D so it is converted to a 2D space using a dimensionality reduction technique. Now our goal is to map the user vectors to the 2D gridworld by minimizing the total distance. It is an NP-hard problem, hence we propose simple greedy algorithm. It is almost same as the traveling salesman problem. After calculating similarities between $n^2$ user vectors and gridworld, the user vectors are mapped to the nearest gridworld point one by one.

\subsection{Learning the \textbf{Q}-Function}\label{sec:training}
  
    In this gridworld environment, the agent is looping over all states and evaluating the $Q$ function for each of the four possible actions. $Q$-learning~\cite{watkins1989learning} and SARSA~\cite{rummery1994line} are frequently used for this problem. We tested both algorithms in this study. Moreover, any state in our environment can be the start state. Then, the policy is updated to select the actions that maximize the $Q$ value at each state. Moreover, we use the $\epsilon$-greedy method for balancing exploration and exploitation~\cite{sutton1998reinforcement}. The policy is described as follows:
    \begin{align}
\pi(s) = \left\{\begin{matrix}
\: \textrm{random action from} \: A& \textrm{if } \xi < \epsilon \\ 
\: \arg\max_{a \in A} \: Q(s,a) & \textrm{otherwise}
\end{matrix}\right.
    \end{align}
    Then, an optimal policy can be found by the Bellman equation. Executing these updates repeatedly is guaranteed to converge to the optimal policy~\cite{sutton1998reinforcement}. In other words, this corresponds to actions that guide the user to obtain good recommendations, while maximizing rewards.

\subsection{Generating Recommendations}\label{sec:generating_recommendations}

    \SetAlgoNoLine
    \begin{algorithm}[t]
        \SetKwInOut{Input}{input}\SetKwInOut{Output}{output}
            \Input{ state-space $S$, action-space $A$, policy $\pi$, transition function $T$, \# candidate start states $k$, a user}
            \Output{recommended the list of items}
        \BlankLine
        select top-$k$ states $\{s_1,..,s_k\}$ with high similarity to a user;\\
        
        \For{ $i \leftarrow 1 \; \KwTo \; k$ }{
            $s \leftarrow s_i$;\\
            \While{at least one items to recommend}{
                recommend items based on $s$;\\
                $a \leftarrow$ $\epsilon$-greedy with $\pi(s)$;\\
                execute action $a$;\\
                $s' \leftarrow T(s,a)$;\\
                $s \leftarrow s'$;\\
            }
        }
        \caption{{\sc Generating Recommendations}}\label{alg:gen_recommendations}
    \end{algorithm}\DecMargin{1em}
    
    Algorithm \ref{alg:gen_recommendations} describes a procedure for generating recommendations for a user. Unlike the other tabula methods, all the states in this environment can be the starting state. To set the starting state, the jaccard distance between all states and the user is calculated. The state with the highest similarity to the user becomes the starting state. Then, the algorithm attempts to recommend the item with the $\epsilon$-greedy based policy until there are no more recommended items.
 
\subsection{Updating the Model Online}
    One of the major advantages of the proposed model is that the user feedback is reflected the states online. This makes the value of the reward function different. As the value of the reward function changes, the optimal policy may change. For example, when a user who has recommended an itemset in state $s_t$ is satisfied with the items, the system immediately adds that user to the userset $U_{s_t}$ of the corresponding state $s_t$. If the user are satisfied with the item recommended in the next state $s_{t+1}$, the size of $U_{s_{t+1}}$ is also increased by 1. As a result, $ \textrm{R}(s_t,a_t,s_{t+1}) $ increases from $|U_{s_t} \cap U_{s_{t+1}}| \: / \: |U_{s_t} \cup U_{s_{t+1}}|$ to $|U_{s_t} \cap U_{s_{t+1}}|+1 \: / \: |U_{s_t} \cup U_{s_{t+1}}|+1$. Using the algorithm proposed in this paper, it is possible to update the state space in real time, and since the reward value changes according to the updated state space, the recommendation can be changed according to the current trend of the users.
   
\section{Experiments}\label{sec:experiments}
\subsection{Dataset}
    OpenAI Gym~\cite{openai} is used to experiment the proposed algorithm. Gym has a collection of environments so that the proposed reinforcement learning can be easily implemented. In addition, two Movielens datasets are used for evaluating our algorithm. One is a dataset of 943 users and 1682 items including 100,000 ratings. The other one has 1,000,209 ratings with 6,040 users and 3,900 items. Both data sets represent the preferences of users as ratings from 1 to 5. Two Movielens datasets were binarized applying a threshold of three rating points, as done in many studies. 80\% of the dataset was used as the training set and the remaining was used to test the algorithm. To evaluate the algorithms in the cold start conditions, each test user rating was deleted leaving only 10\%.

\subsection{States Setup}
    To obtain the biclusters from the matrix, we use two well-known biclustering algorithms: Bimax~\cite{prelic2006systematic}, Bibit~\cite{rodriguez2011biclustering}. These two biclustering algorithms have the minimum number of rows and columns of the biclusters as input parameter. We vary the parameter values and obtain biclusters of various shapes and sizes. Subsequently, we randomly select a total of $n^2$ biclusters to map to the gridworld. Finally, the state space in the gridworld is completed using the proposed greedy algorithm. In our experiment, $n$ is 20 and 30 on the two Movielens datasets so that the total number of states is 400 and 900, respectively.
    
\subsection{Q-Learning versus SARSA}
    \begin{figure}[t]
    	\begin{center}
    		\includegraphics[width=3.3in]{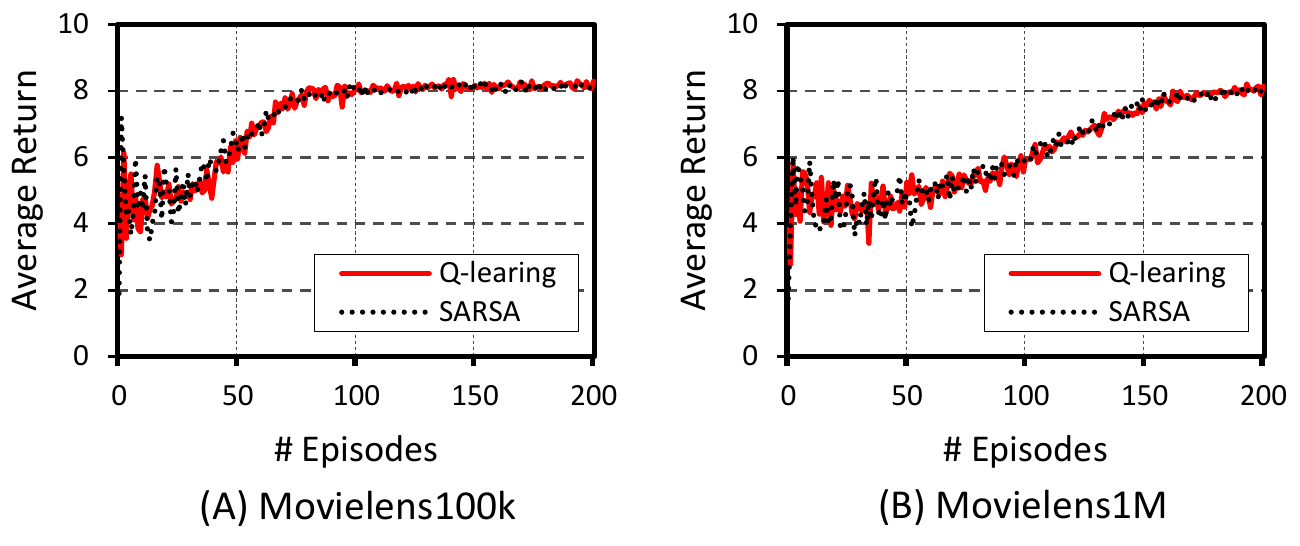}
    	\end{center}
    	\caption{\small Learning curve on Movielens dataset}
    	\label{fig:learning_curve}
    \end{figure}
	Q-learning and SARSA are used to find the optimal policy $\pi$. Q-learning is an off-policy based method while SARSA is an on-policy based method. In this paper, we used both methods and evaluated which performed better in our environment. Figure \ref{fig:learning_curve} demonstrates the learning curve on the two Movielens datasets. The average return is measured by the number of episodes. The performance of the two algorithms are almost identical in the experiment.
    
\subsection{Performance on the Cold-Start Problem}
    \ctable[
        caption = {P@30 and R@30 Comparison},
        label = tab:rec_evaluation,
        % doinside = \scriptsize,
        mincapwidth = 3.2in,
        pos=t
    ]{lcccc}{
    }{
    	\toprule
    	& \multicolumn{2}{c}{\textbf{Movielens\_100k}} & \multicolumn{2}{c}{\textbf{Movielens\_1M}}\\ \cline{2-5}
    	& P@30 & R@30 & P@30 & R@30\\
    	\midrule 
        Global-average & 0.153 & 0.102 & 0.161 & 0.094 \\
        User-based & 0.187 & 0.129 & 0.212 & 0.119 \\
        Item-based & 0.193 & 0.132 & 0.220 & 0.124 \\
        Proposed & \textbf{0.246} & \textbf{0.169} & \textbf{0.277} & \textbf{0.155}\\
        \bottomrule
    }
    
    In this paper, we use ranking metrics to evaluate proposed algorithm. The two most popular ranking metrics are precision and recall. Given a top-N recommendation list $I_N$, precision and recall are defined as follows:
    
    \begin{align} 
    \textrm{P@N} = \frac{|I_N \cap I_{D}|}{N} \\
    \textrm{R@N} = \frac{|I_N \cap I_{D}|}{|I_{D}|} 
    \end{align}
    where $I_{D}$ is the items that algorithm should predict. The precision and recall are calculated by averaging the precision and recall over all the users.
    
	We evaluate the standard algorithms based on which of the items are actually hidden by the user in the test data. Table \ref{tab:rec_evaluation} shows the results for P@30 and R@30 on the two Movielens datasets. The general observation is that the proposed algorithm outperforms the other recommendation methods under the cold-start condition. 

\subsection{Explainable Recommendation}
    When recommending an item group with this proposed algorithm, the reason for the recommendation can be explained to users by informing the corresponding state $s$ together~\eg{ group of items $I_s$ or group of users $U_s$}. Providing a reason for a recommendation to a user in a real system can increase the recommendation reliability. In a model-based based recommender system, the recommendations are not explainable. In addition, existing bicluster based recommendation systems are not usually suitable for providing explanations because the bicluster size is usually very large.

\section{Discussion}\label{sec:dicussion}
  In this paper, we assume that $n^2$ good quality of biclusters are given.
However, biclustering is heuristically found in most cases, due to the fact that it is an NP-hard problem. Therefore, the performance related to providing a recommendation depends largely on the biclusters. We leave finding an optimal bicluster as a part of future research.
    Moreover, we have reduced the action space to four, top, bottom, right and left, but this action space can move in 8 directions or more directions in a multi-dimensional space instead of 2D space. It can be easily extended. Of course, the larger the action space, the greater the computational complexity and the better the accuracy of the recommendation. The method of mapping the bicluster to the state space is crucial for the quality of the recommendation. In addition, we are unable to test with various evaluation methods such as coverage, novelty, etc., but it is expected that the value of coverage and novelty will be acceptable based on $\epsilon$ value.

%\balance

\section{Conclusion}\label{sec:conclusion}
	We have proposed a novel algorithm using RL and biclustering to mitigate the cold-start, online-learning, and explainable recommendation problems. We formulate a recommender system as a gridworld game by using a biclustering technique that reduces the state and action space significantly. Using biclustering not only reduce the space but also improves the recommendation quality of the cold start problem. Moreover, the system can explain to users why the system recommends certain items. We examine the proposed algorithm on the real world dataset and achieved better performance than standard recommender technique. We expect that this algorithm will be useful for recommending items in actual commercial applications.
    
\section*{Acknowedgement}
This work was supported in part by a research grant from Naver Corporation and in part by Samsung Research Funding Center of Samsung Electronics under Project Number SRFC-IT1601-05.
% [Recommender system based on deep learning and information theory] (partially) funded by Naver corporation.

\bibliographystyle{ACM-Reference-Format}
\bibliography{sigproc} 

\end{document}